\documentclass[12pt, oneside]{article}   	
\usepackage[margin=1 in ]{geometry}                		
\usepackage{graphicx}				
\usepackage{cite}
\usepackage{algorithm}
\usepackage[noend]{algpseudocode}

\usepackage{cite}
\usepackage{amssymb,amsfonts}
\usepackage{algpseudocode}
\usepackage{multicol}
\usepackage{graphicx}
\usepackage{textcomp}
\usepackage{xcolor}
\usepackage{flushend}
\usepackage{amsmath}
\usepackage{amssymb}
\usepackage{amsfonts}
\usepackage{float}
\usepackage{graphicx}
\usepackage{fancyref}
\usepackage{booktabs}
\usepackage{subfigure}

\graphicspath{ {./Compressed Sensing/} }

\usepackage{bm}
\usepackage{tikz}
\usetikzlibrary{shapes.geometric, arrows}
\tikzstyle{startstop} = [rectangle, rounded corners, minimum width=3cm, minimum height=1cm,text centered, draw=black, fill=red!30]
\tikzstyle{io} = [trapezium, trapezium left angle=70, trapezium right angle=110, minimum width=3cm, minimum height=1cm, text centered, draw=black, fill=blue!30]
\tikzstyle{process} = [rectangle, minimum width=3cm, minimum height=1cm, text centered, draw=black, fill=orange!30]
\tikzstyle{decision} = [diamond, minimum width=3cm, minimum height=1cm, text centered, draw=black, fill=green!30]
\tikzstyle{arrow} = [thick,->,>=stealth]

\title{\vspace{-2 cm} \textbf{Projection assisted Dynamic Mode Decomposition of large scale data}}
\author{\textbf{Mohammad N. Murshed, M. Monir Uddin}\\ Department of Mathematics and Physics\\North South University\\ Dhaka, Bangladesh}

\usepackage{subfigure}
\usepackage{graphicx}

\newlength\figureheight 
\newlength\figurewidth

\makeatletter
\def\algbackskip{\hskip-\ALG@thistlm}
\makeatother

\usepackage{amsthm}
 
\theoremstyle{definition}
\newtheorem{definition}{Definition}[section]

\begin{document}
\maketitle

\section*{\centering Abstract}
We have deluge of data in time series format for numerous phenomena. The number of snapshots, resolution and many other factors come into play as we look to identify the dynamics in a given problem. The pre-processing and post-processing steps while working with the data may be related to the resources in terms of the hardware used to collect data and computations that we perform on the data to create a model for the problem. Dynamic Mode Decomposition (DMD) is a data based modeling tool that identifies a matrix to map a quantity at some time instant to the same quantity in future. It is possible to generate a model by projecting the high dimensional spatiotemporal data to a lower dimensional subspace in a probabilistic framework. Sampling and gaussian projection have been used in the past to increase efficiency in the computation. Here, we design an optimized version of DMD that utilizes time delay coordinates and a projection matrix. In our proposal, we discussed about two projection matrices -- one is inspired by the Krylov subspace and the other promotes and leverages sparsity to bring computational benefits in producing a model. Satisfactory results are obtained as they are tested on data related to Double gyre (present in ocean mixing) and on a 2D compressible signal. The motivation behind this scheme of DMD comes from the fact that data from many phenomena are 'big' and 'highly oscillatory.' \\ \\
\textbf{Keywords:} projection matrix, time delay coordinates, dynamic mode decomposition, krylov subspace, sparse random matrix, double gyre, compressible signal.


\section{Introduction}
Dynamic Mode Decomposition, \cite{schmid2010dynamic}, is a data-informed modeling technique that has been in use since 2007. If we have some spatiotemporal data available for a phenomenon from any field like fluid dynamics, stock market or epidemiology, then DMD can extract the dynamics from the data and predict the states in the future. This equation free method is able to construct an approximate model for most of the problems, but need to be updated for a few of them. For instance, DMD would have to use time delay coordinates in case the data is highly oscillatory. Incorporation of time delay coordinates make DMD more robust and usable for many problems around us. There are many versions of DMD designed to suit different kinds of problems.\\

Often, the data is high dimensional and processing all of them in the DMD algorithm may not be a good idea. Compressing the high dimensional data to lower dimensional space yields a matrix that will save both memory and time as DMD processes to come up with a model. Such procedure is called Compressed DMD, \cite{brunton2013compressive}, by the practitioners. Although, this is the focus of this work, it is necessary to note that there is another variant of DMD known as compressive sampling DMD that takes advantage of the fact that signals or images (compressible) can be sampled at very low rate (Candes, Romberg, Tao and Donoho) and reconstructed via \(l_1\) minimization. \\ 

Compressing data requires projection matrices (random matrices that are uniformly or normally distributed). A clever way to reduce floating point operations would be to make the projection matrices sparse.\\

 
Sampling and projections have been used as a pre-processing step in DMD to minimize computational burden. The data sequence to be used as the input into the DMD algorithm is represented by a small matrix known as sketch. Illustrated in \cite{brunton2013compressive,kutz2016dynamic} is how compressive sampling theories can be applied to the input matrix to get a compressed version of the available data to efficiently produce a model. Tu proposed a compressed DMD routine and got promising results after its application on a compressible signal, \cite{jtu}. Authors in \cite{erichson2019randomized} deploy the tool on sea surface temperature (SST) data. The foreground and background parts in a video can be separated by compressed DMD, \cite{erichson2019compressed}. Randomized DMD is also exploited to develop reduced order models for complex fluid flows, \cite{bistrian2017randomized}. \\ 

\begin{figure}
\centering
\begin{tikzpicture}[node distance=2cm]
\node (start)[startstop]{Data};
\node (pro2a) [process, below of=start, yshift=-0.5cm] {Sparse Projected Data};
\node (pro2b) [process, right of=start, xshift=2cm] {Sampled Data};
\node (dec1) [decision, below of=pro2b, yshift=-0.5cm] {TDC-DMD};
\node (out1) [io, right of=dec1,xshift=3.5 cm] {Modes/ Spectrum};
\draw [arrow] (start) -- (pro2a);
\draw [arrow] (start) -- (pro2b);
\draw [arrow] (start) -- (pro2b);
\draw [arrow] (pro2a) --(dec1);
\draw [arrow] (pro2b) --(dec1);
\draw [arrow] (dec1) --(out1);
\end{tikzpicture}\\
\caption{Architecture of Sparse Random Projection enabled TDC-DMD}
\label{fig:Arch}
\end{figure}
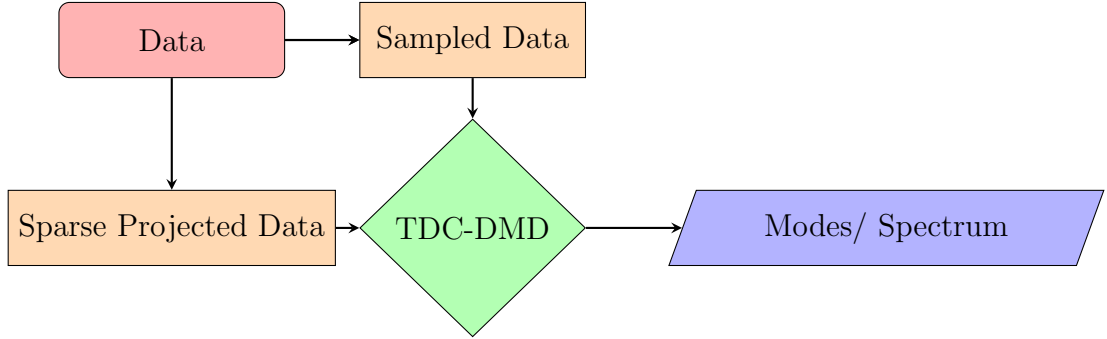

We provide, in this work, a thorough analysis of the mechanics of the randomized DMD and test different projections such as sparse random projection and Krylov subspace to come up with a model for the Double Gyre and 2D compressible signal. The subtleties are explained in terms of probability and statistics. The full state data is both sampled at certain number of spatial points and projected via Random projection. Afterwards, time delay coordinate based DMD (TDC-DMD) is to be applied on these low dimensional data to derive the dynamic modes and eigenvalue spectrum (Figure \ref{fig:Arch}). We assess the performance of each variant of randomized DMD for future prediction.
\\ \\
The rest of the paper is organized as follows: Section \ref{background} explains the idea of DMD and its time delay coordinate version. Krylov Subspace method is also clarified in the same section. Afterwards, projection assisted TDC-DMD, which is the main contribution of this article, is discussed in Section \ref{RDMD}. Section \ref{PF} formulates the two problems under consideration. The results of deploying this new version of TDC-DMD is then shown and compared to that from applying classical TDC-DMD, sampling based TDC-DMD and Gaussian projection based TDC-DMD. 
\\ \\
In this paper, we have used the following notations: '\(\dagger\)' indicates pseudoinverse, '*' refers to the conjugate transpose, '\~{}' symbolizes the low rank version of a matrix. 
\section{Background}
\label{background}

         This section presents the theories, definitions and algorithms necessary to develop Randomized DMD. The reader may take a look at \cite{trefethen1997numerical} to review some of the basic concepts of linear algebra (if necessary). The dynamic mode decomposition and time delay coordinates are discussed. Then, we briefly go over why sampling comes handy and how it is done. 

\subsection{Dynamic Mode Decomposition}
Let's assume that we have some spatiotemporal data, \(\textbf{X} \in \mathbb{R}^{M \times N}\),
\[\textbf{X}=\begin{bmatrix}
    \textbf{x}_{11} & \textbf{x}_{12} &  ... & \textbf{x}_{1N} \\
    \textbf{x}_{21} & \textbf{x}_{22} &  ... & \textbf{x}_{2N} \\
    \textbf{x}_{31} & \textbf{x}_{32} &... & \textbf{x}_{3N} \\
     \vdots            & \vdots              &  \ddots & \textbf{x}_{(M-1)N} \\
    \textbf{x}_{M1} & \textbf{x}_{M2} &  ... & \textbf{x}_{MN} \\
    \end{bmatrix},
\]
for a problem in the form of a matrix where the columns represent a certain quantity for different spatial coordinates at a particular time instant. \(M\) is the number of spatial nodes and \(N\) the number of temporal nodes. Dynamic Mode Decomposition splits this data sequence into two parts, \(\textbf{X}_1^{N-1}\) and \(\textbf{X}_2^{N}\), and runs according to Algorithm \ref{ALGO:1} to output a model to predict the future. To better clarify, for \(M=5\) and \(N=3\), the two parts will be,
\[
\textbf{X}_1^{2}=\begin{bmatrix}
    \textbf{x}_{11} & \textbf{x}_{12} \\
    \textbf{x}_{21} & \textbf{x}_{22} \\
    \textbf{x}_{31} & \textbf{x}_{32} \\
    \textbf{x}_{41} & \textbf{x}_{42} \\
    \textbf{x}_{51} & \textbf{x}_{52} \\
    \end{bmatrix}, \
 \textbf{X}_2^{3}=\begin{bmatrix}
    \textbf{x}_{12} &  \textbf{x}_{13} \\
    \textbf{x}_{22} &  \textbf{x}_{23} \\
    \textbf{x}_{32} &   \textbf{x}_{33} \\
    \textbf{x}_{42} &   \textbf{x}_{43} \\
    \textbf{x}_{52} &   \textbf{x}_{53} \\
    \end{bmatrix}.
\]
The output from DMD will then be the solution at the fourth temporal node and so on upto a certain temporal node we are interested about. DMD originates from the Koopman operator, an infinite dimensional linear operator to map the current states to future states.\\

\begin{algorithm}
    \caption{DMD}\label{ALGO:1}
    \hspace*{\algorithmicindent} \textbf{Input}: $\textbf{X}_1^{N-1}, \textbf{X}_2^{N}$ \\
    \hspace*{\algorithmicindent} \textbf{Output}: $\textbf{x}_{DMD}(t)$
    \begin{algorithmic}[1]
    \State $ \textbf{X}_1^{N-1}=\textbf{U}\Sigma \textbf{V}^{*}$
    \State $ r \ = \ rank(\textbf{X}_{1}^{N-1}) $
    \State $ \textbf{U}=\textbf{U} [ \ :  \ , \ : r ] $
    \State $ \Sigma= \Sigma [ \ :  r , \ : r ] $
    \State $ \textbf{V}=\textbf{V} [ \ :  \ , \ : r ] $
    \State $\tilde{S} = \textbf{U}^{*}\textbf{X}_2^{N} \textbf{V} \Sigma^{-1}$
    \State $\tilde{S} \textbf{y}_{k} = \mu_{k} \textbf{y}_k$
    \State $\phi_{k} = \textbf{U} \textbf{y}_{k}$.
    \State $\omega_{k}= \frac{ln(\mu_{k})} {\Delta t} \) and \(\textbf{b}=\Phi^{\dagger}\textbf{x}_1$.
    \State $\textbf{x}_{DMD}(t) =  \sum_{k=1}^{r} b_{k}(0) \phi_{k}(\textbf{x}) exp(\omega t) = \Phi \ \ diag(exp(\omega t)) \textbf{b} $.
    \end{algorithmic}
    \end{algorithm}


\noindent \textbf{Connection to Koopman Operator and Proper Orthogonal Decomposition}\\ \\
DMD is a special case of Koopman Mode Decomposition (KMD), \cite{kutz2016dynamic}. The key difference is that DMD works on state space whereas KMD takes place in observable space:
$$ g(\textbf{x}_{k+1})=\textbf{K}g(\textbf{x}_k). $$ 
An interesting aspect in DMD is the singular value decomposition of the first data sequence, \(\textbf{X}_1^{N-1}\), that results in \(\textbf{U}\). The columns of \(\textbf{U}\) are called proper orthogonal decomposition modes. The idea behind Proper Orthogonal Decomposition (POD), \cite{sirovich1987turbulence, chatterjee2000introduction}, is to use snapshots to represent a flow, \(f(x,t)\), using time coefficients, \(h(t)\), and spatial modes, \(g(x)\), as
\begin{equation}
f(x,t)=h(t) g(x).
\end{equation}
As the name suggests, POD searches for as few (optimal) orthogonal bases as possible to approximate the flow. More information on the concept and applications of POD can be found in \cite{taira2017modal}. Next, we reveal the critical points of DMD and a few situations where DMD fails to learn from data.\\

\noindent \textbf{Limitations of Dynamic Mode Decomposition}\\ \\
\textbf{Sampling Strategy}. Usually, the data fed into DMD are collected based on a certain frequency. For DMD to work properly, it is necessary that sampling is enacted at exactly equal to or greater than twice the maximum frequency in the data flow (signal). \\ \\
\textbf{Zero vector initial condition.} Note that in the coordinate transformation step  \(\textbf{b}=\Phi^{\dagger}\textbf{x}_1\), \textbf{b} turns out to be a zero vector in case \(\textbf{x}_1\) is a zero vector (zero inital condition). \\ \\
\textbf{Highly oscillatory data.} DMD fails to correctly identify the frequencies when the data contain oscillations. In the following section, we elaborate on a way to circumvent these two issues. 

\subsection{Time delay coordinate based DMD}
Time delay coordinate based DMD, \cite{kutz2016dynamic}, is a variant of DMD that uses augmented data sequence (in the form of Hankel matrices) to generate a model. The dataset is augmented vertically by adding a copy of the time shifted states. Consider the state vectors for 6 temporal nodes 
\(\textbf{x}_{1}, \textbf{x}_{2}, \textbf{x}_{3}, 
\textbf{x}_{4}, \textbf{x}_{5}, \textbf{x}_{6}\). We can 
construct \(\textbf{X}_{1, aug}\) and \(\textbf{X}_{2, aug}\) (before running DMD) as
\[
\textbf{X}_{1, aug}=\begin{bmatrix}
    \textbf{x}_{1} \ \textbf{x}_{2} \  \textbf{x}_{3} \ \textbf{x}_{4}\\
    \textbf{x}_{2} \ \textbf{x}_{3} \ \textbf{x}_{4} \ \textbf{x}_{5}\\
    \end{bmatrix}, \ 
\textbf{X}_{2, aug}=\begin{bmatrix}
    \textbf{x}_{2} \ \textbf{x}_{3} \  \textbf{x}_{4} \ \textbf{x}_{5}\\
    \textbf{x}_{3} \ \textbf{x}_{4} \ \textbf{x}_{5} \ \textbf{x}_{6}\\
    \end{bmatrix}.
\]
The routine for time delay coordinate based DMD is provided in Algorithm \ref{ALGO:2}. We define the variable \(q\) as 
$$ q =  1 +  p, $$
where \(p\) is the number of time shifted states. For the instance above, 1 copy of time shifted data is used, hence, \(q = 2\). It is imperative to carefully set the value of \(q\) so to capture the dynamic modes.  This algorithm is essentially classic DMD when \(q=1\). \\


\begin{algorithm}
    \caption{Time Delay Coordinate based DMD}\label{ALGO:2}
    \hspace*{\algorithmicindent} \textbf{Input}: $ \textbf{X}_{1,aug}^{N-1}, \textbf{X}_{2,aug}^{N}$ \\
    \hspace*{\algorithmicindent} \textbf{Output}: $\textbf{x}_{DMD}(t)$
    \begin{algorithmic}[1]
    \State $ \textbf{X}_{1,aug}^{N-1}=\textbf{U}\Sigma \textbf{V}^{*}$
    \State $ r \ = \ rank(\textbf{X}_{1,aug}^{N-1}) $
    \State $ \textbf{U}=\textbf{U} [ \ :  \ , \ : r ] $
    \State $ \Sigma= \Sigma [ \ :  r , \ : r ] $
    \State $ \textbf{V}=\textbf{V} [ \ :  \ , \ : r ] $
    \State $\tilde{S} = \textbf{U}^{*}\textbf{X}_{2,aug}^{N} \textbf{V} \Sigma^{-1}$
    \State $\tilde{S} \textbf{y}_{k} = \mu_{k} \textbf{y}_k$
    \State $\phi_{k} = \textbf{U} \textbf{y}_{k}$.
    \State $\omega_{k}= \frac{ln(\mu_{k})} {\Delta t} \) and \(\textbf{b}=\Phi^{\dagger}\textbf{x}_1$.
    \State $\textbf{x}_{DMD}(t) =  \sum_{k=1}^{r} b_{k}(0) \phi_{k}(\textbf{x}) exp(\omega t) = \Phi \ \ diag(exp(\omega t)) \textbf{b} $.
    \State $\textbf{x}_{DMD}(t) =  \textbf{x}_{DMD}(t) [ \ :  M, \ :  N ] $.
    \end{algorithmic}
    \end{algorithm}

The large size of the augmented matrices reduces computational efficiency if \(M \gg N\). Random sampling aids in bypassing this problem. Sampling means to collect a subset of the complete signal. There are several different ways in statistics to sample from a population. Here, we deploy random sampling without replacement to identify a subset of the all the spatial nodes in the domain. This is much like placing sensor on a few locations rather than using sensors at all the spatial nodes in the domain. Although, fewer measurements are expected, it is instructive to have enough measurements to retain the structure of the original signal at a given a spatial coordinate.

\subsection{Krylov subspace with Arnoldi methods}
Recently, Krylov subspace based projection method is considered as
an efficient tool for solving may mathematical problems. Here, we apply
this method to perform the projection based time delay coordinate DMD (TDC-DMD).
\begin{definition}[Krylov subspace]\label{krylovsubspace} 
Consider $A\in \mathbb{R}^{n\times n}$, $b \in \mathbb{R}^n$ and
 a set of linearly independent vectors, \index{linearly independent}
 \begin{equation} \label{eq:reviewofla:krylovbasis}
 V_m =\{b, Ab,\cdots, A^{m-1}b \}.
 \end{equation}
A subspace which is spanned by $V_m$ is called $m$ dimensional
\emph{Krylov subspace} associated with $A$ and $b$ and
it can be defined as
\begin{equation}\label{eq:reviewofla:krylovsubspace}
\mathcal{K}_m(A,b)= span \{V_m\}.
\end{equation}
\end{definition}
To construct a basis of Krylov subspace,
Arnoldi process is considered as one of the
efficient methods.\\


The explicit formulation of Krylov basis
$V_m$ in (\ref{eq:reviewofla:krylovbasis})
is not suitable for numerical computations. As $m$
increases, the vector $A^mb$ always converges to an eigenvector
belonging to a dominant eigenvalue.
This implies that  the vectors in $V_m$ become more and more linearly dependent. To avoid
these effects, one should choose a basis of a better nature, for
example an orthonormal basis. However, in this case we can follow the modified Gram-Schmidt
procedure  introduced above.
Exploiting the idea of modified Gram-Schmidt procedure
to form the orthonormal basis of Krylov subspace is known as
Arnoldi procedure. In fact the Arnoldi procedure applies the Gram-Schmidt procedure
to transform the vectors in $V_m$ into orthonormal set
of vectors $\{v_1, v_2,\cdots, v_m\}$ which form a basis
of $m$-dimensional Krylov subspace $\mathcal{K}_m(A,b)$.
The Arnoldi procedure  is summarized
in Algorithm~\ref{alg:reviewofla:arnoldi4krylovbasis}.

 
 \begin{algorithm}
    \caption{Arnoldi process for orthonormal basis of Krylov subspace.}
    \label{alg:reviewofla:arnoldi4krylovbasis}
    \hspace*{\algorithmicindent} \textbf{Input}: $A\in\mathbb{R}^{n\times n}, b\in{\mathbb{R}^n}$. \\
    \hspace*{\algorithmicindent} \textbf{Output}: Orthogonal set of vectors $v_1,~v_2,~\cdots,~ v_m$.
    \begin{algorithmic}[1]
     \State $v_1=\frac{b}{\|b\|_2}$ 
    \For{$i=1:m$}
    \State $w=Av_i$.\;  
            \For{$j=1:i$}
                \State $h=v_j^Tw$.\;
                \State $w=w-hv_j$.
             \EndFor   
             \If{$\|w\|_2\leq \epsilon$ (a tolerance)}
                \State {stop.}
             \Else
                \State $v_i=\frac{w}{\|w\|_2}$.
              \EndIf
      \EndFor      
    \end{algorithmic}
    \end{algorithm}

\section{Projection based time delay coordinate Dynamic Mode Decomposition} 
\label{RDMD}
This section outlines the optimized version of TDC-DMD we came up with. We utilize the time delay coordinate based DMD and propose two projection methods -- one employs Krylov Subspace and the other relies heavily on sparsity. Time delay coordinates can handle high oscillation and projection matrices reduce the computational stress due to the large size of the data sequence.

\subsection{Random Projection}
It is a technique \cite{bingham2001random} to transform a higher dimensional matrix to a lower dimensional one by the use of a random matrix \((R)\). The equation to move into a lower dimensional subspace \((a \ll M)\) will read:
\begin{equation}
\textbf{X}_{a \times N}= \textbf{R}_{a \times M} \textbf{X}_{M \times N}.
\end{equation}
The \(R\) matrix is in general orthogonal. The gram matrix must be or somewhat close to identity matrix i.e.
\begin{equation}
\textbf{R}^{*} \textbf{R} \approx \textbf{I}. 
\end{equation}
Random matrices used previously with DMD include the gaussian random matrix and the uniform random matrix. For instance, uniform/gaussian projections and single pixel measurement were applied to identify the dynamic modes for sparse linear dynamics in the Fourier domain in\cite{kutz2016dynamic}. We introduce two new projections: the first one is produced from Arnoldi vectors and second one depends on sparsity. Either of these is then incorporated into the TDC-DMD to design an efficient version of TDC-DMD, the steps of which are delineated in Algorithm \ref{PDMD}. 

\subsection{Arnoldi vectors for projection}
Arnoldi iteration is an iterative method, based on modified Gram-Schmidt orthogonalisation, that takes a random matrix and an arbitrary vector to generate a matrix (\textbf{V}) containing orthonormal bases and an upper triangular matrix. The orthononormal vectors in \(\textbf{V}\) are called Arnoldi vectors. Taking inner product of the Arnoldi vectors yields 0, that is (for \(i \neq j\)),
\begin{equation}
\textbf{v}_{i} . \textbf{v}_{j} = 0.
\end{equation}
We use a random matrix of dimension \(M \times M\), a vector of ones of dimension \(M \times 1\) and the value of \(a\) (\(a \ll M\)) for the dimension of the Krylov Subspace as input to the Algorithm \ref{alg:reviewofla:arnoldi4krylovbasis}. The resulting orthonormal matrix is then transposed and multiplied by the data sequence to project the high dimensional data to lower dimensional subspace:
\begin{equation}
\textbf{X}_{(a+1) \times N}= \textbf{V}^{*}_{(a+1) \times M} \textbf{X}_{M \times N}. 
\end{equation}
Note that the \(h\) in Algorithm \ref{alg:reviewofla:arnoldi4krylovbasis} forms a Hessenberg matrix which is built upon the Arnoldi vectors:
\[\textbf{\~{H}}=\begin{bmatrix}
    h_{1,1} & h_{1,2} &  h_{1,3} & \hdots & h_{1,N} \\
    h_{2,1} & h_{2,2} & h_{2,3} &  \hdots & h_{2,N} \\
    0 & h_{3,2} & h_{3,3} &  \hdots & h_{3,N} \\
     \vdots            & \ddots              &  \ddots & \ddots & \vdots \\
     \vdots            & \vdots              &  0 & h_{N-1,N} & h_{N,N} \\
    0 & \hdots & \hdots & 0 & h_{N+1, N} \\
    \end{bmatrix}.
\]
This whole idea is often used to efficiently solve eigenvalue problems.

\subsection{Sparse Random Projection}
We use the Achlioptas random matrix as the sparse random projector to reduce computational stress in data processing. \\

\textbf{Achlioptas Random Matrix.} Achlioptas \cite{achlioptas2001database} proposed a random projection matrix with entries based on the distribution below with \(s\) being either 1 or 3:

\[ R_{ij}=\sqrt{s}\begin{cases} 

     -1 & \text{ with  probability $1/(2s)$}\\

      0 & \text{ with  probability $1-1/s$} \\

      1 & \text{ with  probability $1/(2s)$} 

   \end{cases}.
\]

Such matrix allows processing of just a fraction of the complete high dimensional data, thereby saving memory and time. 


\begin{algorithm}
    \caption{Projection enabled DMD}\label{PDMD}
    \hspace*{\algorithmicindent} \textbf{Input}: $ \textbf{R}, \textbf{X}_{1,aug}^{N-1}, \textbf{X}_{2,aug}^{N}$ \\
    \hspace*{\algorithmicindent} \textbf{Output}: $\textbf{x}_{DMD}(t)$
    \begin{algorithmic}[1]
    \State $ \textbf{Z}_{1,aug}^{N-1}= \textbf{R}\textbf{X}_{1,aug}^{N-1},  \textbf{Z}_{2,aug}^{N}= \textbf{R}\textbf{X}_{2,aug}^{N} $
    \State $ \textbf{Z}_{1,aug}^{N-1}=\textbf{U}\Sigma \textbf{V}^{*}$
    \State $ r \ = \ rank(\textbf{Z}_{1,aug}^{N-1}) $
    \State $ \textbf{U}=\textbf{U} [ \ :  \ , \ : r ] $
    \State $ \Sigma= \Sigma [ \ :  r , \ : r ] $
    \State $ \textbf{V}=\textbf{V} [ \ :  \ , \ : r ] $
    \State $ \bm{A = U^{*}} \textbf{Z}_{2,aug}^{N} \bm{{V \Sigma}}^{-1} $
    \State $ \bm{A W= W \Sigma} $
    \State $ \bm{\Phi_{X}} = \textbf{X}_{2,aug} \bm{V_{Z} \Sigma_{Z}^{-1} W_{Z}}  $
    \State $\omega_{k}= \frac{ln(\mu_{k})} {\Delta t} \) and \(\textbf{b}=\Phi^{\dagger}\textbf{x}_1$.
    \State $\textbf{x}_{DMD}(t) = \bm{\Phi_{X}} \ \ diag(exp(\bm{\omega} t)) \textbf{b} $.
    \State $\textbf{x}_{DMD}(t) =  \textbf{x}_{DMD}(t) [ \ :  M, \ :  N ] $.
    \end{algorithmic}
    \end{algorithm}

\section{Numerical Results} 
\label{NR}
The classical TDC-DMD, sample-based TDC-DMD, and projection enabled TDC-DMD are applied on the problem of Double Gyre and 2D compressible signal to extract the dynamic modes and create a model for prediction. All the results are produced using Python 3.5 (on an Intel CORE i5 processor with 8 GB 1600 MHz DDR3 memory).

\subsection{Double Gyre}\label{PF}
\begin{figure}
\centering \includegraphics[scale=0.6]{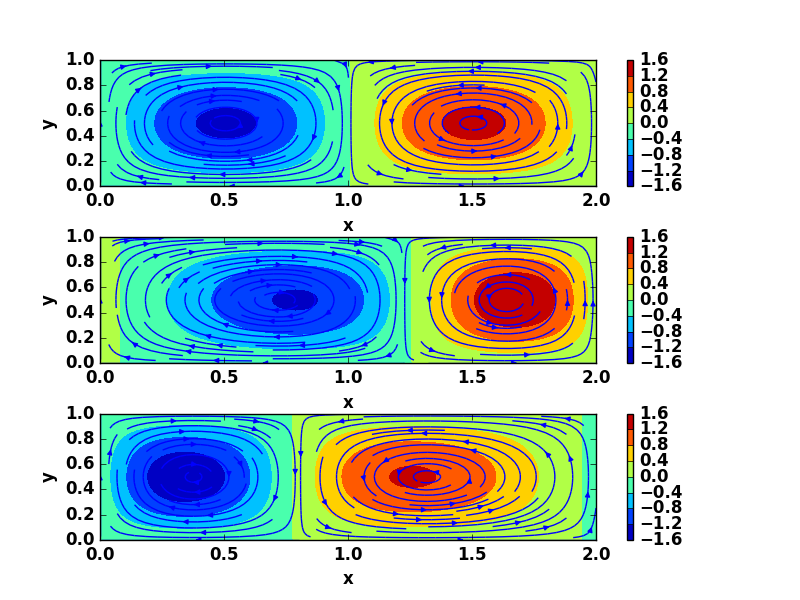}
\caption{Double Gyre Dynamics for t = 0.05 (top), t = 2.00 (middle), t = 6.50 (bottom)}
\label{fig:DoubGyre}
\end{figure}

A gyre is a system of circulating currents that form due to Coriolis effect. This phenomenon is typically a result of the wind motion through the landmass as the Earth is rotating. A double gyre is an incompressible flow where two counter rotating vortices expand and contract periodically, Figure \ref{fig:DoubGyre}. Its model is defined by the stream-function: 
\begin{equation}
 \Psi(x,y,t) = A sin( \pi f(x,t)) sin ( \pi y),
 \end{equation}
 where 
\begin{equation}
f(x,t) = \epsilon sin(\omega t) x^{2} +x -2 \epsilon sin(\omega t) x .
\end{equation}
Note that \((x,y)\) refers to the spatial coordinates and \(t\) the time coordinate.
The horizontal and vertical velocities are derived from the spatial derivatives of the streamfunctions on the domain [0,2] \(\times\) [0,1]: 
\begin{equation}
u= -\frac{\partial \Psi}{\partial y}=\pi A sin(\pi f(x)) cos(\pi y),
\end{equation}
\begin{equation}
v=\frac{\partial \Psi}{\partial x}= \pi A cos (\pi f(x) ) sin( \pi y) \frac{df}{dx}.
\end{equation}
Parameters used are \(A= 0.1\), \(\omega= 2 \pi/10\),\(\epsilon=0.25\). The vorticity, computed from the spatial derivatives of the velocities, will read as
\begin{equation}
vorticity= \frac{\partial v}{\partial x} -\frac{\partial u}{\partial y}.
\end{equation}

\begin{center}
Table 1: Number of measurements for different versions of DMD\\
\begin{tabular}{llr}
\toprule
{} &                   Variant &  Measurements \\
\midrule
0 &                   Classic &         10000 \\
1 &                  Sampling &           100 \\
2 &       Gaussian Projection &           200 \\
3 &  Sparse Random Projection &           100 \\
4 &       Krylov Matrix based &           100 \\
\bottomrule
\end{tabular}

\end{center}

\begin{figure}
\centering  
\subfigure[Classical]{\includegraphics[scale=0.4]{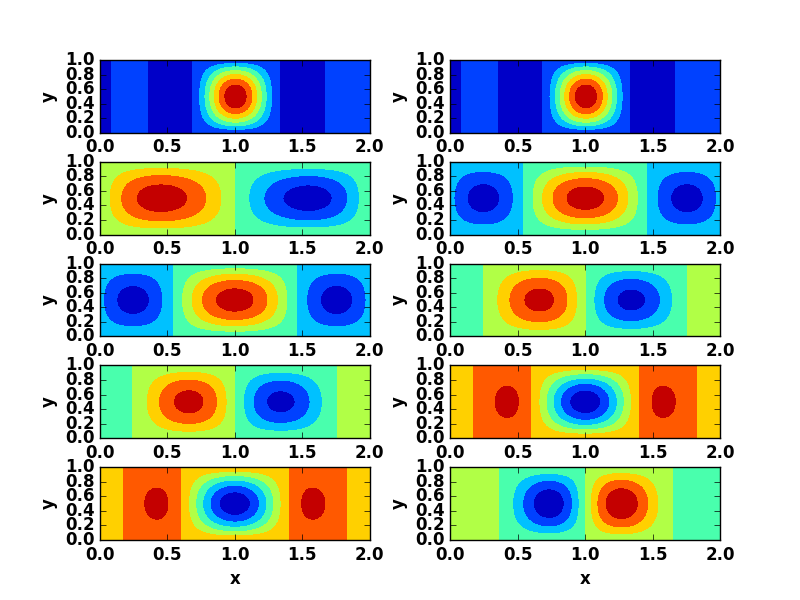}\label{DMD_C}}
\subfigure[POD modes]{\includegraphics[scale=0.4]{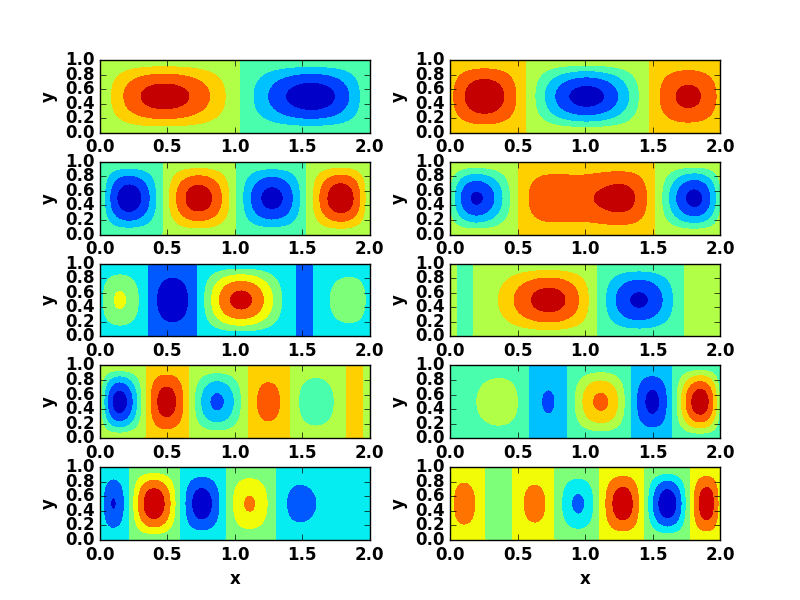}\label{POD_m}}
\caption{Classical DMD modes and POD modes}
\end{figure}
\begin{figure}
\centering
\subfigure[Sample]{\includegraphics[scale=0.4]{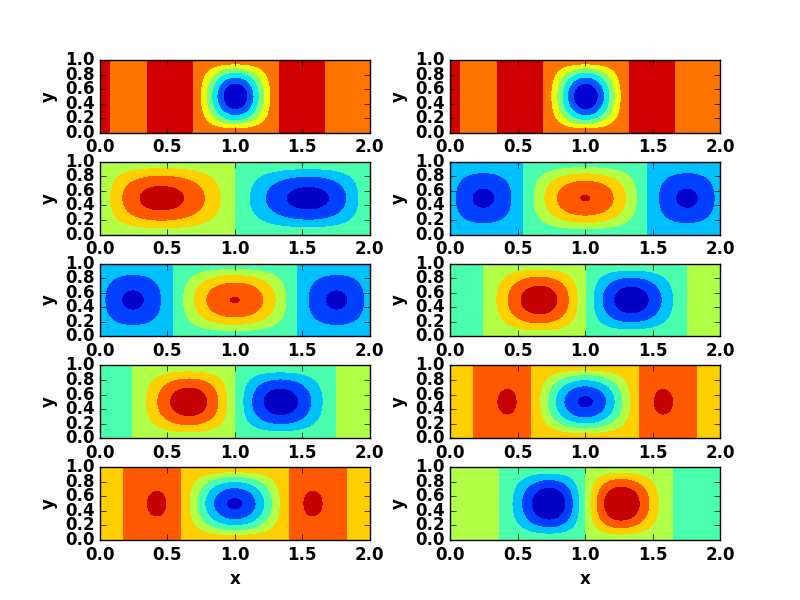}}
\subfigure[Gauss Projection]{\includegraphics[scale=0.4]{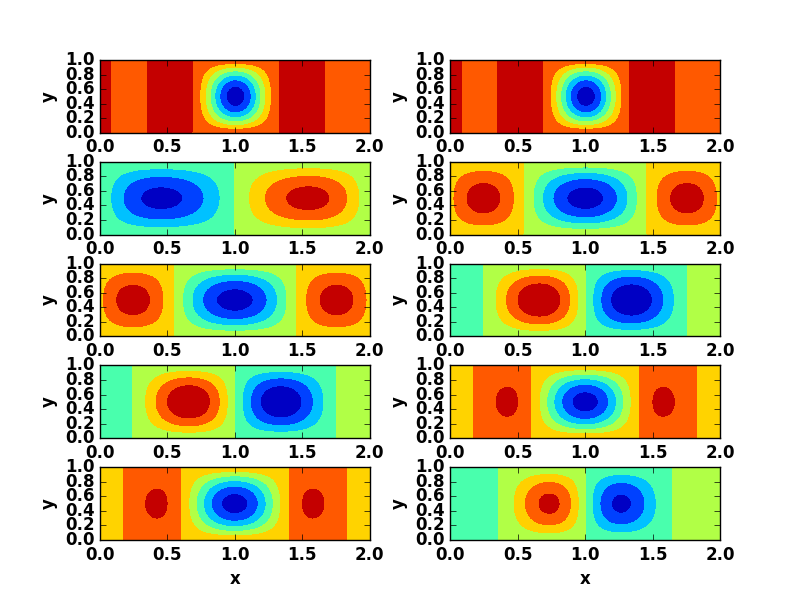}}
\subfigure[Arnoldi vector]{\includegraphics[scale=0.4]{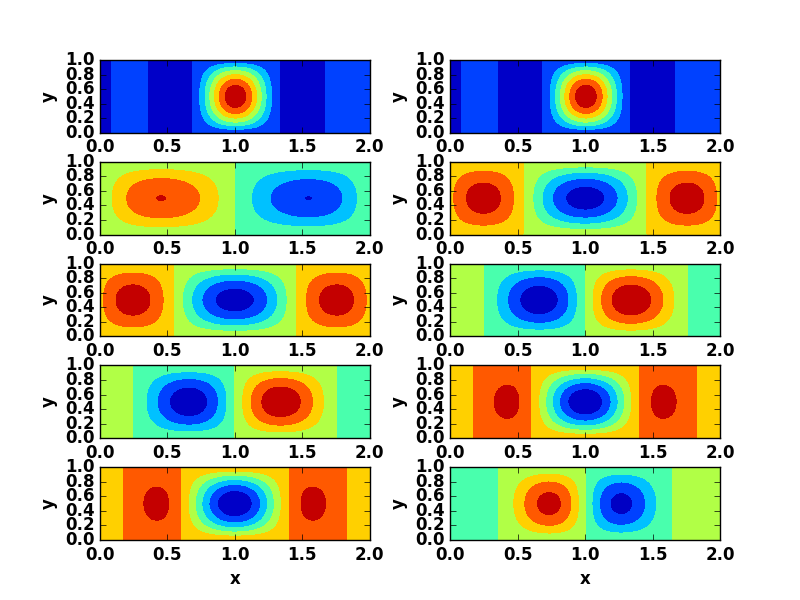}}
\subfigure[Sparse Random Projection]{\includegraphics[scale=0.4]{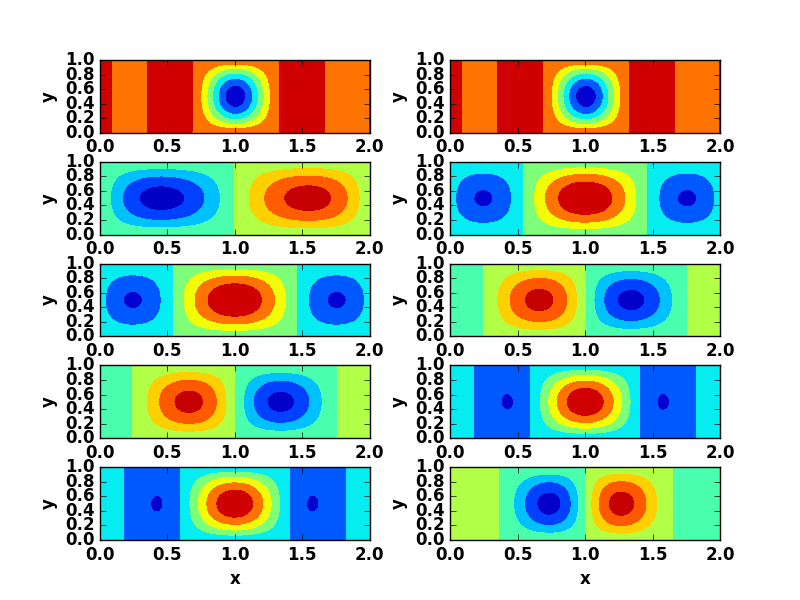}}
\caption{Four variants of DMD}
\label{DMD_OV}
\end{figure}

\begin{figure}
\centering  
\subfigure[Spectrum]{\includegraphics[scale=0.45]{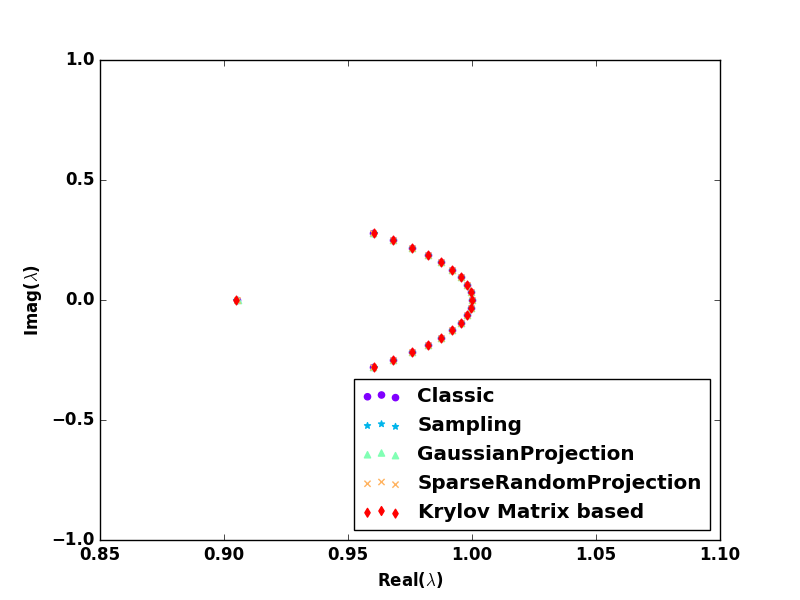} \label{SPEC}}
\subfigure[Error Analysis]{\includegraphics[scale=0.45]{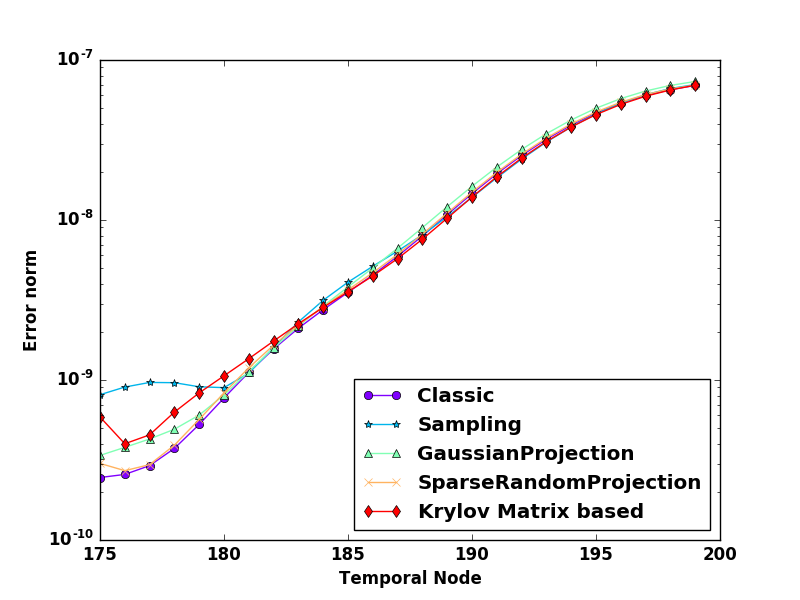} \label{EA}}
\caption{DMD Spectrum and Long run error analysis}
\end{figure}

The Double Gyre is simulated for around 10 s by the use of 200 temporal nodes. The vorticity field for the first 174 temporal nodes are used as training data and the rest to check the validity of the DMD model. The complete code is at first run to identify the number of dominant modes in the data. We have used the first 20 singular values to form the low rank mapping matrix.\\

The number of measurements for each variant of DMD are given in Table 1. Sampling and projection enabled DMD use a small percentage (approximately 1 to 2 percent) of what is used for the classical DMD. It is important to start with an arbitrary number of measurements and then keep on adjusting till we get a reliable model. \\

Depicted in Figure \ref{POD_m} are the POD modes. They appear to be a blend of multiple different modes. The dynamic modes, found from the optimized variants of DMD, are illustrated in Figure \ref{DMD_OV} and that from the traditional DMD in Figure \ref{DMD_C}. The discrepancies could be due to the difference in the input data for these four variants. The dimension of the input matrix varies for each variant resulting in different singular values as singular value decomposition (SVD) is performed.\\

For real-valued data, the eigenvalues are symmetric about \(Im(\lambda)=0\). Eigenvalues within or on a unit circle imply a faithful model whereas eigenvalues outside of the unit circle is a sign that the model may contain error in the long run. The symmetry in Figure \ref{SPEC} is in agreement with the fact that we have used real data. All the four variants result in the same set of eigenvalues.\\

The model created by Sparse Random Projection enabled DMD performs as good as the models from other variants, Figure \ref{EA}, as long as the number of rows in \(\textbf{X}_{Aug}\) is well above the low rank, \(r\), used for the mapping matrix i.e. \(aq \ge r\), where \(a\) is the number of measurements.  

\subsection{Compressible Signal}
A signal that consists of a few dominant 
frequencies is known as a compressible signal. We use a 2D signal of this type from \cite{jtu} for this example, where gaussians are used for the spatial modes:

\begin{equation}\label{sig}
f(t)= sin(2 \pi f_{1} t) v_{1} + sin(2 \pi f_{2} t) v_{2} + 0.1 w(t)
\end{equation}
\begin{equation}\label{v1}
v_{1}=2 exp ( -\frac{(x-0.5)^2}{2(0.6)^2} - \frac{(y-0.5)^2}{2(0.2)^2})
\end{equation}
\begin{equation}\label{v2}
v_{2}= exp ( -\frac{(x+0.25)^2}{2(0.6)^2} - \frac{(y-0.35)^2}{2(1.2)^2})
\end{equation}
 
\begin{figure}
\centering  
\subfigure[\(v_2\)]{\includegraphics[scale=0.4]{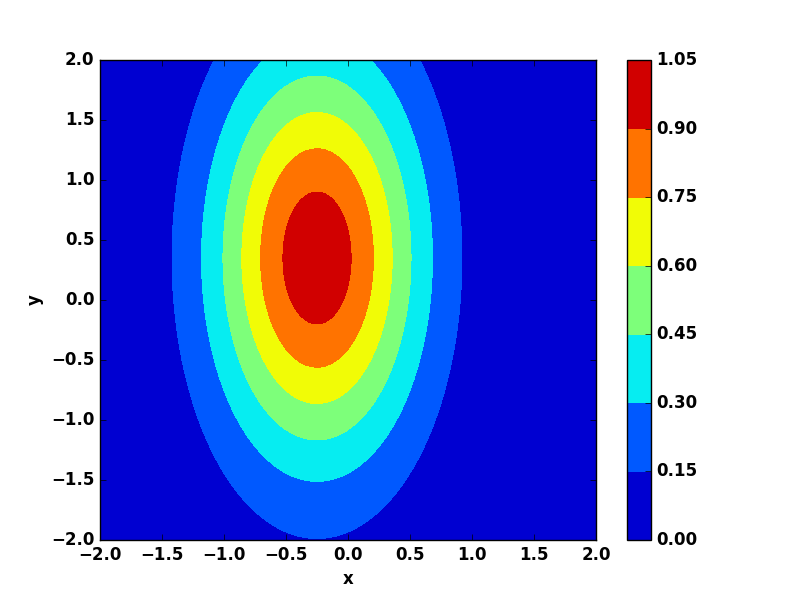}}
\subfigure[\(v_1\)]{\includegraphics[scale=0.4]{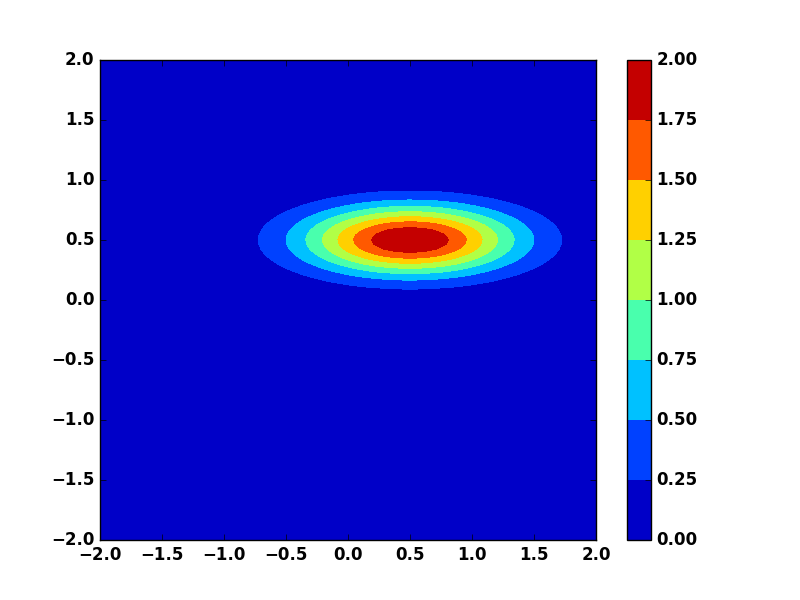}}
\caption{True spatial modes}
\label{Vs}
\end{figure}

\begin{center}
Table 2: Number of measurements for different versions of DMD
\begin{tabular}{llr}
\toprule
{} &                   Variant &  Measurements \\
\midrule
0 &                   Classic &         10000 \\
1 &                  Sampling &           100 \\
2 &       Gaussian Projection &            50 \\
3 &  Sparse Random Projection &            50 \\
4 &       Krylov Matrix based &            50 \\
\bottomrule
\end{tabular}

\end{center}

A nominal signal, (\ref{sig}), with \(f_{1}=1.3\) and \(f_{2}= 8.4\) is generated for \( -2 \le x,y \le2\). The simulation is run with \(\Delta t =0.05\) for \(T \approx 4\). The time step must be adjusted as per the maximum frequency present in the signal.  Any background noise can be incorporated through \(w(t)\). We have set \(w(t)\) to 0 in our example. The spatial modes, (\ref{v1}) and (\ref{v2}), are shown in Fig. \ref{Vs}.\\

\begin{figure}
\centering  
\subfigure[Classical]{\includegraphics[scale=0.4]{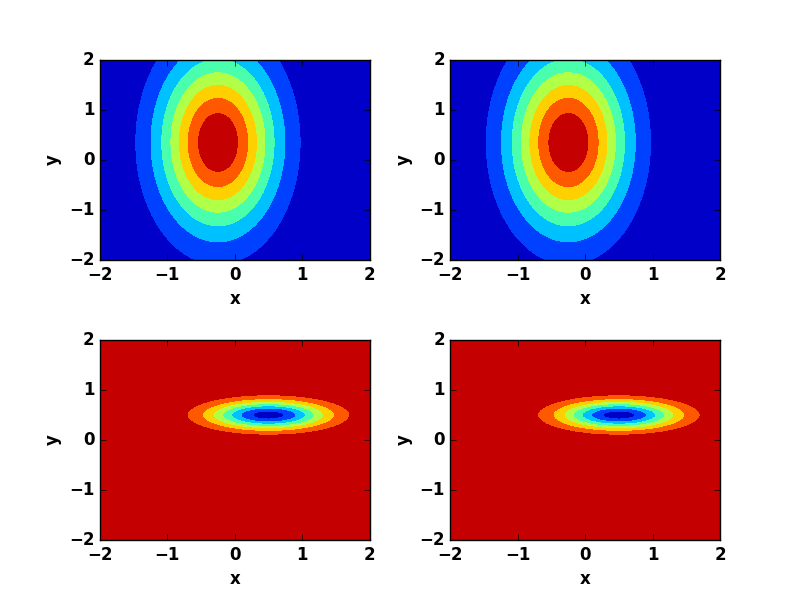}\label{DMD_C}}
\subfigure[POD modes]{\includegraphics[scale=0.4]{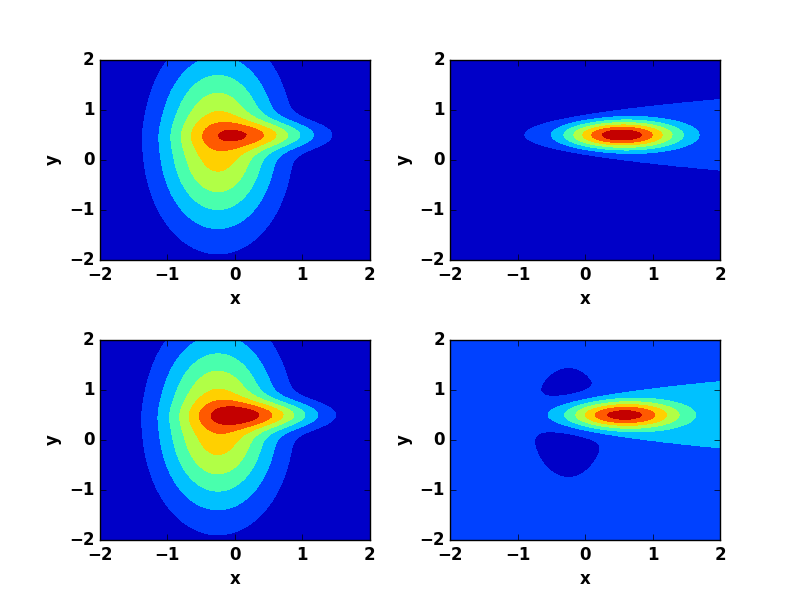}\label{POD_m}}
\caption{Classical DMD modes and POD modes}
\end{figure}
\begin{figure}
\centering
\subfigure[Sample]{\includegraphics[scale=0.4]{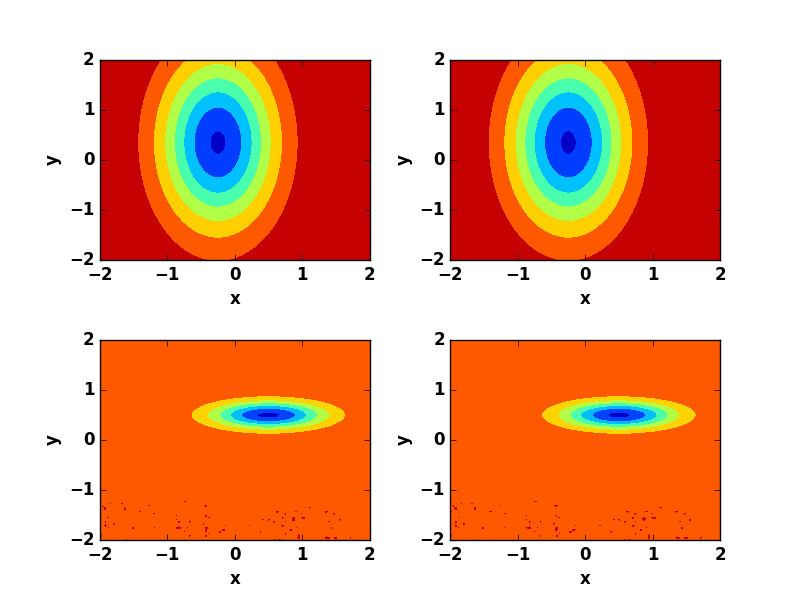}}
\subfigure[Gauss Projection]{\includegraphics[scale=0.4]{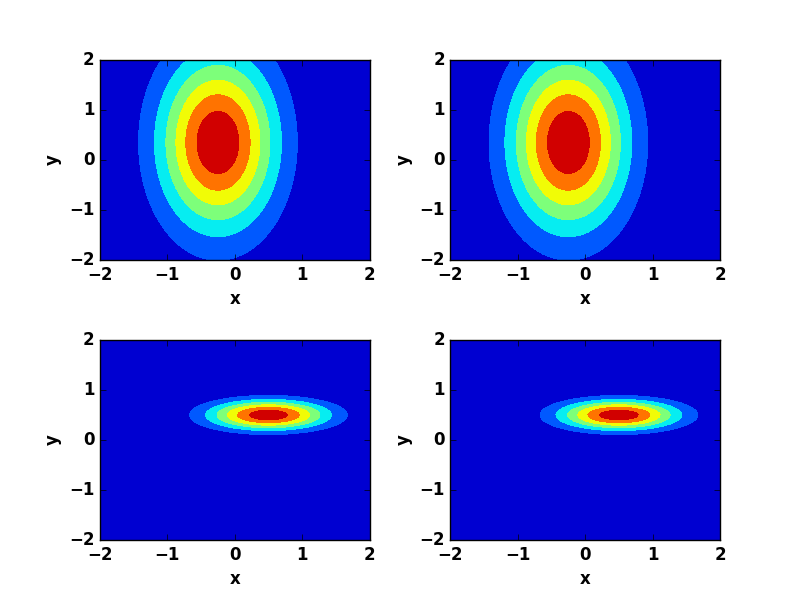}}
\subfigure[Arnoldi vector]{\includegraphics[scale=0.4]{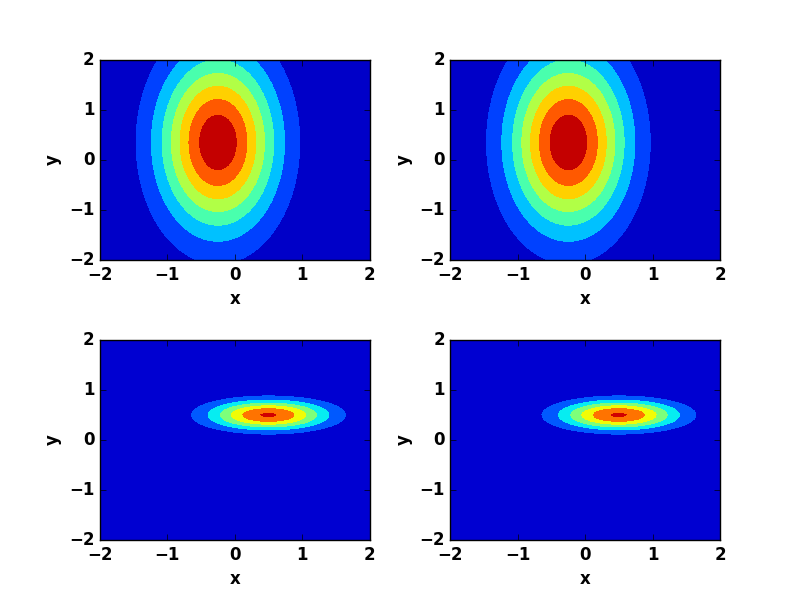}}
\subfigure[Sparse Random Projection]{\includegraphics[scale=0.4]{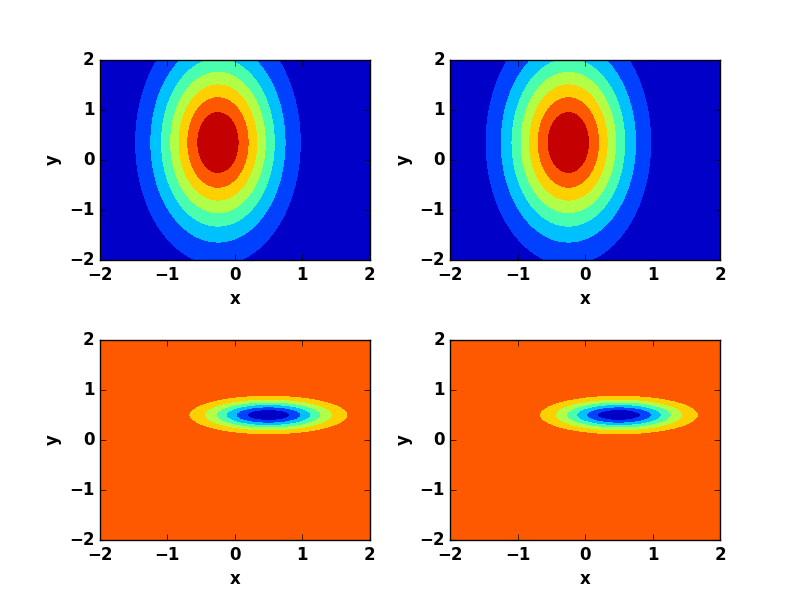}}
\caption{Four variants of DMD}
\label{DMD_OV}
\end{figure}

\begin{figure}
\centering  
\subfigure[Spectrum]{\includegraphics[scale=0.45]{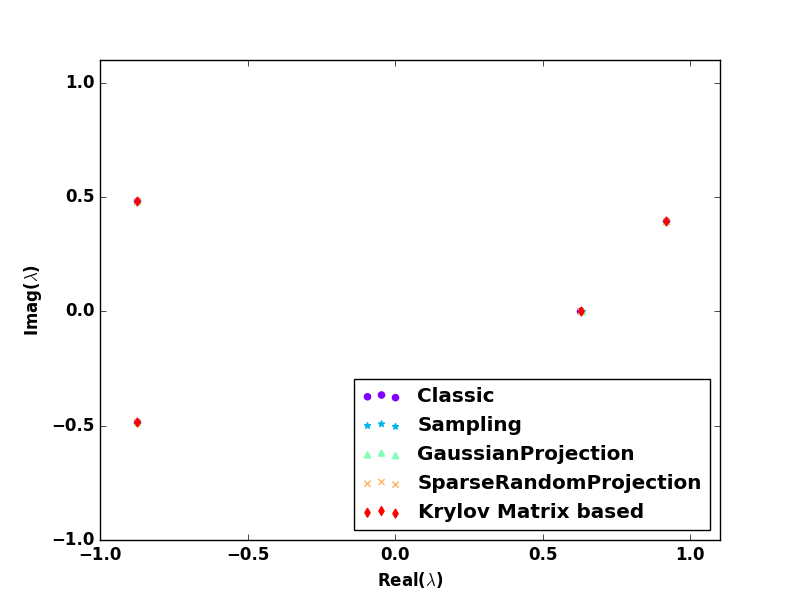} \label{SPEC}}
\subfigure[Error Analysis]{\includegraphics[scale=0.45]{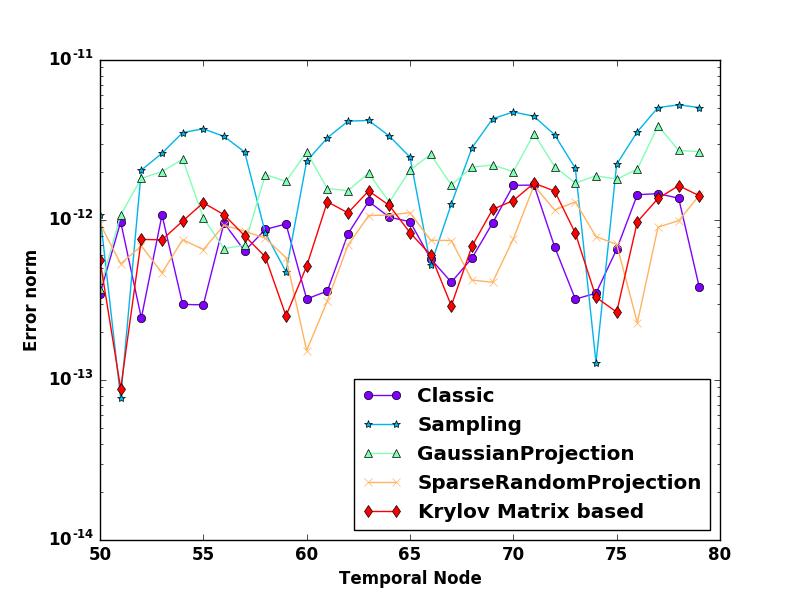} \label{EA}}
\caption{DMD Spectrum and Long Run Error Analysis}
\end{figure}

It is important to note that \(s=3\) results in a reasonable model for this instance. The spectrum in Fig \ref{SPEC} shows that the eigenvalues from all the variants overlap. The dynamic modes are correctly identified by the time delay coordinate based DMD, Fig \ref{DMD_C}. Modes get mixed in proper orthogonal decomposition, Fig. \ref{POD_m}. The dynamic modes from four different variants of the projection assisted time delay coordinate DMD agree well. The number of measurements used in each variant is summarized in Table 2. The error plot, Fig. \ref{EA}, is evidence of satisfactory performance of the Krylov Subspace and Sparse Random projection assisted DMD compared to traditional time delay coordinated DMD that rely on sampling and gaussian projection.


\section*{\centering Conclusion and Future Work}
This paper surveys different variants of TDC-DMD that leverages random projection matrix to reduce the size of the data matrix used as input to the algorithm. All the variants show great promise in generating an approximate model by using a very small percentage of the complete data-set. The Arnoldi vectors from Arnoldi iteration and the Achliptas projection as Sparse Random Projector are powerful tools to bring down the dimension of the data sequence. We can, therefore, apply these projection assisted TDC-DMD on problems with very high-dimensional data, especially when the number of spatial nodes \(\ge 10^4\). A future direction is to compare our results to that from Compressive Sampling DMD which operates in a different way than these projection enabled methods.       

\section*{\centering Acknowledgement}
This project is partially funded by Office of Research, North South University. (Grant number: \textbf{CTRG-19/SEPS/06})
    
\bibliography{Bib_paper_DG}

\begin{thebibliography}{10}
\providecommand{\url}[1]{#1}
\csname url@samestyle\endcsname
\providecommand{\newblock}{\relax}
\providecommand{\bibinfo}[2]{#2}
\providecommand{\BIBentrySTDinterwordspacing}{\spaceskip=0pt\relax}
\providecommand{\BIBentryALTinterwordstretchfactor}{4}
\providecommand{\BIBentryALTinterwordspacing}{\spaceskip=\fontdimen2\font plus
\BIBentryALTinterwordstretchfactor\fontdimen3\font minus
  \fontdimen4\font\relax}
\providecommand{\BIBforeignlanguage}[2]{{%
\expandafter\ifx\csname l@#1\endcsname\relax
\typeout{** WARNING: IEEEtran.bst: No hyphenation pattern has been}%
\typeout{** loaded for the language `#1'. Using the pattern for}%
\typeout{** the default language instead.}%
\else
\language=\csname l@#1\endcsname
\fi
#2}}
\providecommand{\BIBdecl}{\relax}
\BIBdecl

\bibitem{schmid2010dynamic}
P.~J. Schmid, ``Dynamic mode decomposition of numerical and experimental
  data,'' \emph{Journal of fluid mechanics}, vol. 656, pp. 5--28, 2010.

\bibitem{brunton2013compressive}
S.~L. Brunton, J.~L. Proctor, and J.~N. Kutz, ``Compressive sampling and
  dynamic mode decomposition,'' \emph{arXiv preprint arXiv:1312.5186}, 2013.

\bibitem{kutz2016dynamic}
J.~N. Kutz, S.~L. Brunton, B.~W. Brunton, and J.~L. Proctor, \emph{Dynamic mode
  decomposition: data-driven modeling of complex systems}.\hskip 1em plus 0.5em
  minus 0.4em\relax SIAM, 2016.

\bibitem{jtu}
J.~H. Tu, ``Dynamic mode decomposition: Theory and applications,'' Ph.D.
  dissertation, Princeton University, Princeton, 2013.

\bibitem{erichson2019randomized}
N.~B. Erichson, L.~Mathelin, J.~N. Kutz, and S.~L. Brunton, ``Randomized
  dynamic mode decomposition,'' \emph{SIAM Journal on Applied Dynamical
  Systems}, vol.~18, no.~4, pp. 1867--1891, 2019.

\bibitem{erichson2019compressed}
N.~B. Erichson, S.~L. Brunton, and J.~N. Kutz, ``Compressed dynamic mode
  decomposition for background modeling,'' \emph{Journal of Real-Time Image
  Processing}, vol.~16, no.~5, pp. 1479--1492, 2019.

\bibitem{bistrian2017randomized}
D.~A. Bistrian and I.~M. Navon, ``Randomized dynamic mode decomposition for
  nonintrusive reduced order modelling,'' \emph{International Journal for
  Numerical Methods in Engineering}, vol. 112, no.~1, pp. 3--25, 2017.

\bibitem{trefethen1997numerical}
L.~N. Trefethen and D.~Bau~III, \emph{Numerical linear algebra}.\hskip 1em plus
  0.5em minus 0.4em\relax Siam, 1997, vol.~50.

\bibitem{sirovich1987turbulence}
L.~Sirovich, ``Turbulence and the dynamics of coherent structures. i. coherent
  structures,'' \emph{Quarterly of applied mathematics}, vol.~45, no.~3, pp.
  561--571, 1987.

\bibitem{chatterjee2000introduction}
A.~Chatterjee, ``An introduction to the proper orthogonal decomposition,''
  \emph{Current science}, pp. 808--817, 2000.

\bibitem{taira2017modal}
K.~Taira, S.~L. Brunton, S.~T. Dawson, C.~W. Rowley, T.~Colonius, B.~J. McKeon,
  O.~T. Schmidt, S.~Gordeyev, V.~Theofilis, and L.~S. Ukeiley, ``Modal analysis
  of fluid flows: An overview,'' \emph{Aiaa Journal}, pp. 4013--4041, 2017.

\bibitem{bingham2001random}
E.~Bingham and H.~Mannila, ``Random projection in dimensionality reduction:
  applications to image and text data,'' in \emph{Proceedings of the seventh
  ACM SIGKDD international conference on Knowledge discovery and data
  mining}.\hskip 1em plus 0.5em minus 0.4em\relax ACM, 2001, pp. 245--250.

\bibitem{achlioptas2001database}
D.~Achlioptas, ``Database-friendly random projections,'' in \emph{Proceedings
  of the twentieth ACM SIGMOD-SIGACT-SIGART symposium on Principles of database
  systems}.\hskip 1em plus 0.5em minus 0.4em\relax ACM, 2001, pp. 274--281.

\end{thebibliography}
\bibliographystyle{IEEEtran} 

\end{document}